\documentclass[9pt,twocolumn,twoside]{osajnl}
\usepackage{gensymb}
\journal{ol} 

\setboolean{shortarticle}{true}

\title{Design of ultracompact broadband focusing spectrometers based on deep diffractive neural networks}

\author[1]{Yilin Zhu}
\author[2]{Yuyao Chen}
\author[1,2,3,*]{Luca Dal Negro}

\affil[1]{Division of Materials Science and Engineering, Boston University, 15 Saint Mary’s Street, Brookline, Massachusetts 02246, USA}
\affil[2]{Department of Electrical and Computer Engineering and Photonics Center, Boston University, 8 Saint Mary’s Street, Boston, Massachusetts}
\affil[3]{Department of Physics, Boston University, 590 Commonwealth Avenue, Boston, Massachusetts 02215, USA}

\affil[*]{Corresponding author: dalnegro@bu.edu}

\begin{abstract}
We propose the inverse design of ultracompact, broadband focusing spectrometers based on adaptive deep diffractive neural networks (a-D$^2$NNs). Specifically, we introduce and characterize two-layer diffractive devices with engineered angular dispersion that focus and steer broadband incident radiation along predefined focal trajectories with desired bandwidth and $5$ nm spectral resolution. Moreover, we systematically study the focusing efficiency of two-layer devices with side length $L=100~\mu\mathrm{m}$ and focal length $f=300~\,\mu\mathrm{m}$ across the visible spectrum and we demonstrate accurate reconstruction of the emission spectrum from a commercial superluminescent diode. The proposed a-D$^2$NNs design method extends the capabilities of efficient multi-focal diffractive optical devices to include single-shot focusing spectrometers with customized focal trajectories for applications to ultracompact multispectral imaging and lensless microscopy.   
\end{abstract}

\setboolean{displaycopyright}{true}

\begin{document}

\maketitle
Compact spectrometers play an important role in many research areas, such as in the analysis of materials, the detection of biological cells, the characterization of light sources, and the rapid determination of chemical species \cite{Demtroeder:2014_Laser}. In particular, the design and fabrication of ultracompact micron-size spectrometers capable to match the microscopic nature of the investigated objects is being extensively investigated \cite{He:1998_EDG_AWG,Janz:2004_EDG_AWG,Ma:2013_EDG,Babin:2009_Holo,Peroz:2012_Holo,Calafiore:2014_Holo,Cheben:2007_AWG,Zou:2020_AWG,Zhu:2017_ChirpedMetalens,Britton:2020_ChirpedAxilens,Chen:2020_Axilens_AO,Chen:2020_Axilens_OL,Britton:2021_DualBand_LPR,Khorasaninejad:2016_MetalensSpec}. There are currently several grating-based ultracompact spectrometer devices, such as Echelle diffractive gratings (EDG) \cite{He:1998_EDG_AWG,Janz:2004_EDG_AWG,Ma:2013_EDG}, on-chip digital planar holographic gratings \cite{Babin:2009_Holo,Peroz:2012_Holo,Calafiore:2014_Holo}, arrayed waveguide gratings (AWGs) \cite{He:1998_EDG_AWG,Janz:2004_EDG_AWG,Cheben:2007_AWG,Zou:2020_AWG}, as well as chirped meta-lenses and phase-modulated diffractive axilens devices \cite{Zhu:2017_ChirpedMetalens,Britton:2020_ChirpedAxilens,Chen:2020_Axilens_AO,Chen:2020_Axilens_OL,Britton:2021_DualBand_LPR,Khorasaninejad:2016_MetalensSpec}. Additionally, designs based on randomly scattering spectrometers \cite{Redding:2013_Random}, chirped filament-array gratings \cite{Rahnama:2020_ChirpedFilamentArrayGrating}, wavelength-selective filters \cite{Emadi:2012_LVOF,Li:2021_SWF}, and resonator-based spectrometers \cite{Sharkawy:2001_ResPhotonicDefect,Nitkowski:2008_ResMicroring,Xia:2011_ResMicrodonut} were also recently proposed. However, these structures face severe challenges when considering on-demand, non-conventional responses, such as customized focal trajectories on a detection plane, due to the limited flexibility of traditional design approaches. Moreover, they require sophisticated nanofabrication schemes to integrate dispersive and focusing elements in an ultracompact manner. 

Recently, the advancement of data-driven machine learning techniques in optics and photonics offered novel opportunities for inverse design \cite{Liu:2018_MLInverseDesign,Ma:2021_MLInverseDesignReview,Liu:2021_MLInverseDesign}. In particular, it led to the development of efficient all-optical deep diffractive neural networks (D$^2$NNs) for the engineering of multi-layered devices that are directly trained, based on the definition of a suitable loss function, using error backpropagation within diffractive layers without the need of training datasets \cite{Lin:2018_D2NN}. The fruitful combination of deep learning methods and  diffractive physics have provided abundant degrees of freedom (DOF) to enable design and prototyping of task-specific, on-demand devices \cite{Luo:2019_D2NNFilter,Veli:2021_D2NNTerahertz,Chen:2022_D2NNDualLens} based on multi-layer diffractive optical elements (DOEs). Since the first demonstration in object classification \cite{Lin:2018_D2NN}, D$^2$NNs have been used for the design of different optical elements, including broadband filters \cite{Luo:2019_D2NNFilter}, terahertz pulse shapers \cite{Veli:2021_D2NNTerahertz}, and dual-band ultracompact focusing lenses \cite{Chen:2022_D2NNDualLens}. 

In this paper, we introduce the design of ultracompact diffractive focusing spectrometers based on D$^2$NN augmented by adaptive training, called adaptive D$^2$NN (a-D$^2$NN) \cite{Chen:2022_D2NNDualLens}. The targeted spectrometers consist of two diffractive phase modulation layers trained to angularly disperse and focus broadband incident light onto different spatial positions that form desired spatial trajectories on the detection plane. We study the focusing efficiency and bandwidths of such devices and examine the mapping from wavelengths to the focal spot positions. We also investigate how discretized phase profiles affect the overall focusing efficiency, guiding future device fabrication. Finally, as a proof-of-concept application we demonstrate the successful reconstruction of the spectrum of a commercial superluminescent diode source. 

\begin{figure}[htbp]
    \centering\includegraphics[width=\linewidth]{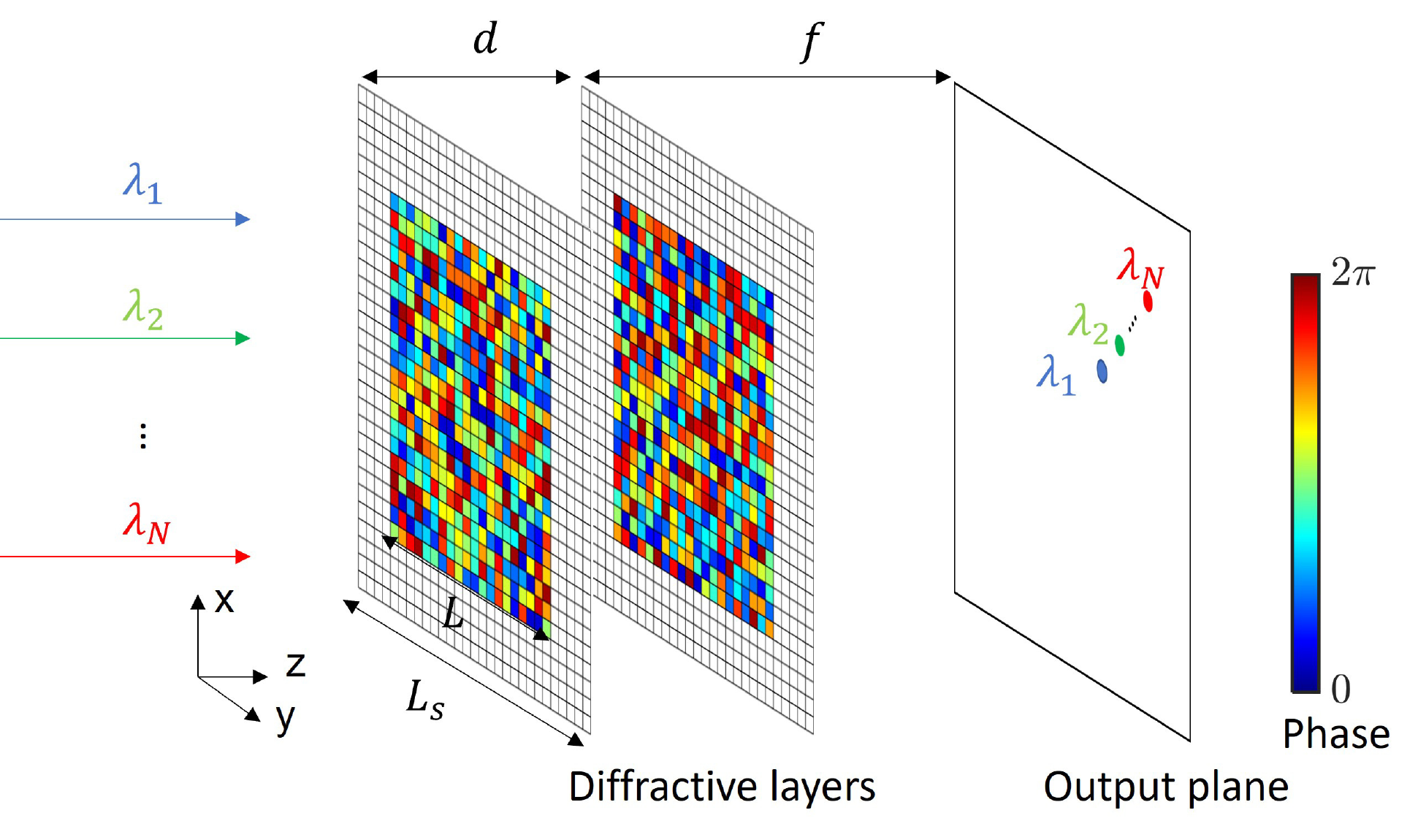}
    \caption{Schematics of inverse design of ultracompact spectrometer device by a-D$^2$NN. The device consists of two diffractive layers and is trained over the phase value of each pixel (phase-only modulation unit). The number of pixels on layers is reduced for visualization purpose. Color coding on diffractive layers represents the phase values. }\label{fig: D2NNschematics}
\end{figure}

In Fig.\,\ref{fig: D2NNschematics} we show the schematics of the designed ultracompact spectrometer consisting of two diffractive square layers acting as phase plates separated by a distance $d$. As a concrete implementation, we considered devices with side length $L=100~\,\mu\mathrm{m}$, inter-layer separation $d=250~\,\mu\mathrm{m}$ and a minimum pixel size $\Delta x=200~\mathrm{nm}$, which can be conveniently fabricated using current diffractive optics and doublet metasurface technology \cite{Banerji:2019_ReviewFlatOptics,Lalanne:2017_MetalensReview,Khorasaninejad:2016_Science,Yilmaz:2019_Metasurface,Britton:2020_ChirpedAxilens,Britton:2021_DualBand_LPR}. The two diffractive layers are discretized into square pixels that modulate the phase of the input wave individually. 
The a-D$^2$NN is trained to optimize the phase modulation $\Delta\phi$ on each pixel in the interval $[0,2\pi)$ with respect to the loss function defined below, while maintaining unit transmittance. Plane waves, with evenly spaced wavelengths $\lambda_1,\,\lambda_2,\,\ldots\lambda_n$ over a broad spectral range, centered at $\lambda_0$, i.e. $\lambda_0=(\lambda_1+\lambda_n)/2$, are used for coherent illumination. These incident waves propagate through the two diffractive layers, then are focused and angularly deflected onto an achromatic output plane at a focal distance $f=300~\,\mu\mathrm{m}$ at desired transverse positions $(x_1,y_1),\,(x_2,y_2),\,\ldots,(x_n,y_n)$. The phase plates are zero-padded around their perimeter in the simulation window ($L_s$ as in Fig.\,\ref{fig: D2NNschematics}) to obtain more accurate diffraction field results from the Rayleigh-Sommerfeld's (RS) first integral formulation \cite{Voelz:2011_MATLAB}: 
\begin{equation} \label{eq: RS integral}
    A_o(x',y')=A_s(x,y)*h(x,y;x',y';z,k)
\end{equation}
\begin{equation} \label{eq: RS kernal}
    h(x,y;z,k)=\frac{1}{2\pi}\frac{z}{r}\left(\frac{1}{r}-jk\right)\frac{e^{jkr}}{r}
\end{equation}
where $r=\sqrt{x^2+y^2+z^2}$ and $z$ stands for the axial distance between $A_o$ and $A_s$ planes. Note that the two phase plates in Fig.\,\ref{fig: D2NNschematics} are located on both sides of a transparent substrate with refractive index $n=3$, as in \cite{Chen:2022_D2NNDualLens}. 

The RS integral given in Eq.\,\ref{eq: RS integral},\,\ref{eq: RS kernal} is used to calculate fields at each layer in the a-D$^2$NN as well as at the output plane. We then introduce the loss function $\mathcal{L}$ defined in terms of the focusing efficiency $\hat{\eta}$. The focusing efficiency $\hat{\eta}$ is the ratio between the power of the focused spot and the one of the incident wave. For each incident wavelength $\lambda_i$ and focal spot position $(x_i,y_i)$, we define $\hat{\eta}$ as \cite{Chen:2022_D2NNDualLens}: 
\begin{equation} \label{eq: focusing efficiency}
    \hat{\eta}(\lambda_i,f;x_i,y_i)=\frac{\int_{0}^{3 \mathrm{FWHM}/2} \mathrm{d}\rho_s^{\prime} \int_{0}^{2 \pi} \mathrm{d}\theta_s^{\prime} ~I^{\prime}\left(\lambda_i, z=d+f, \rho_s^{\prime}, \theta_s^{\prime}\right)}{\iint \mathrm{d}S~I(\lambda_i, z=0, \rho, \theta)}
\end{equation}

Here in Eq.\,\ref{eq: focusing efficiency}, the symbols $(\rho_s',\theta_s')$ denote the polar coordinates on the focal plane with its origin shifted to the point $(x_i,y_i)$. So we have $\rho_s^{\prime}=\sqrt{(x'-x_i)^2+(y'-y_i)^2}$ and $\theta_s^{\prime} = \cos^{-1}[(x^\prime-x_i)/\rho_s^{\prime}]$, where $(x^\prime,y^\prime)$ is the Cartesian coordinates across the focal plane. The denominator simply provides the incident power over the input aperture. The relations above show how the focal position $(x_i,y_i)$ gets incorporated in the efficiency calculation and eventually the loss function $\mathcal{L}$.

According to standard grating theory, the angular dispersion of spectrometer devices is quantified by the resolving power, which is given by \cite{Demtroeder:2014_Laser}: 
\begin{equation} \label{eq: delta lambda}
    L(\mathrm{d}\theta/\mathrm{d}\lambda)=\frac{\lambda_0}{\Delta\lambda}
\end{equation}
$\Delta\lambda$ is the minimum resolvable wavelength (i.e., the spectral resolution) of the spectrometer. We adapt the methodology in \cite{Britton:2020_ChirpedAxilens} to distribute the focal positions $(x_i,y_i)$ at wavelengths $\lambda_i$ along a line with a given direction angle $\alpha$ (with respect to $x$-axis) on the focal plane. Considering the case of linear angular dispersion, the relation between the focal positions and the wavelengths can be obtained as follows:
\begin{equation} \label{eq: focal spot center x}
    x_i=\cos{\alpha}\frac{\lambda_0 f}{L{\Delta}\lambda} (\lambda_i-\lambda_1)+x_1
\end{equation}
\begin{equation} \label{eq: focal spot center y}
    y_i=\sin{\alpha}\frac{\lambda_0 f}{L{\Delta}\lambda} (\lambda_i-\lambda_1)+y_1
\end{equation}

Note that these coordinates are the targeted focal positions at the incident wavelengths, and are incorporated into the loss function $\mathcal{L}$ for its minimization. 

Following this approach, we used the stochastic gradient descent (SGD) method to minimize the loss function in the backpropagation steps. During each epoch, we randomly select $B$\,wavelengths from $\lambda_1,\,\lambda_2,\,\ldots\lambda_n\,(B<n)$ as a mini-batch and feed them into the network. Crucially, in order to improve the convergence, we implemented adaptive loss weights that update along with each loss term during an epoch \cite{Chen:2022_D2NNDualLens}. The loss function for this mini-batch is thus defined as:
\begin{equation}\label{eq: loss funciton}
\mathcal{L} = \sum_{i \in B}w_i\left[1-\hat{\eta}(\lambda_i,f;x_i,y_i)\right]^2
\end{equation}
where $w_i$ is the adaptive loss weight that corresponds to $\lambda_i$. These weights are updated with a coefficient $\gamma$. They are initialized as unity and updated during the $k$-th epoch based on the following rule \cite{Chen:2022_D2NNDualLens}:
\begin{equation}\label{eq: weight update rule}
w_i^k\leftarrow w_i^{k-1}+\gamma\left[1-\hat{\eta}(\lambda_i,f;x_i,y_i)\right]^2
\end{equation}

\begin{figure}[htbp]
    \centering\includegraphics[width=\linewidth]{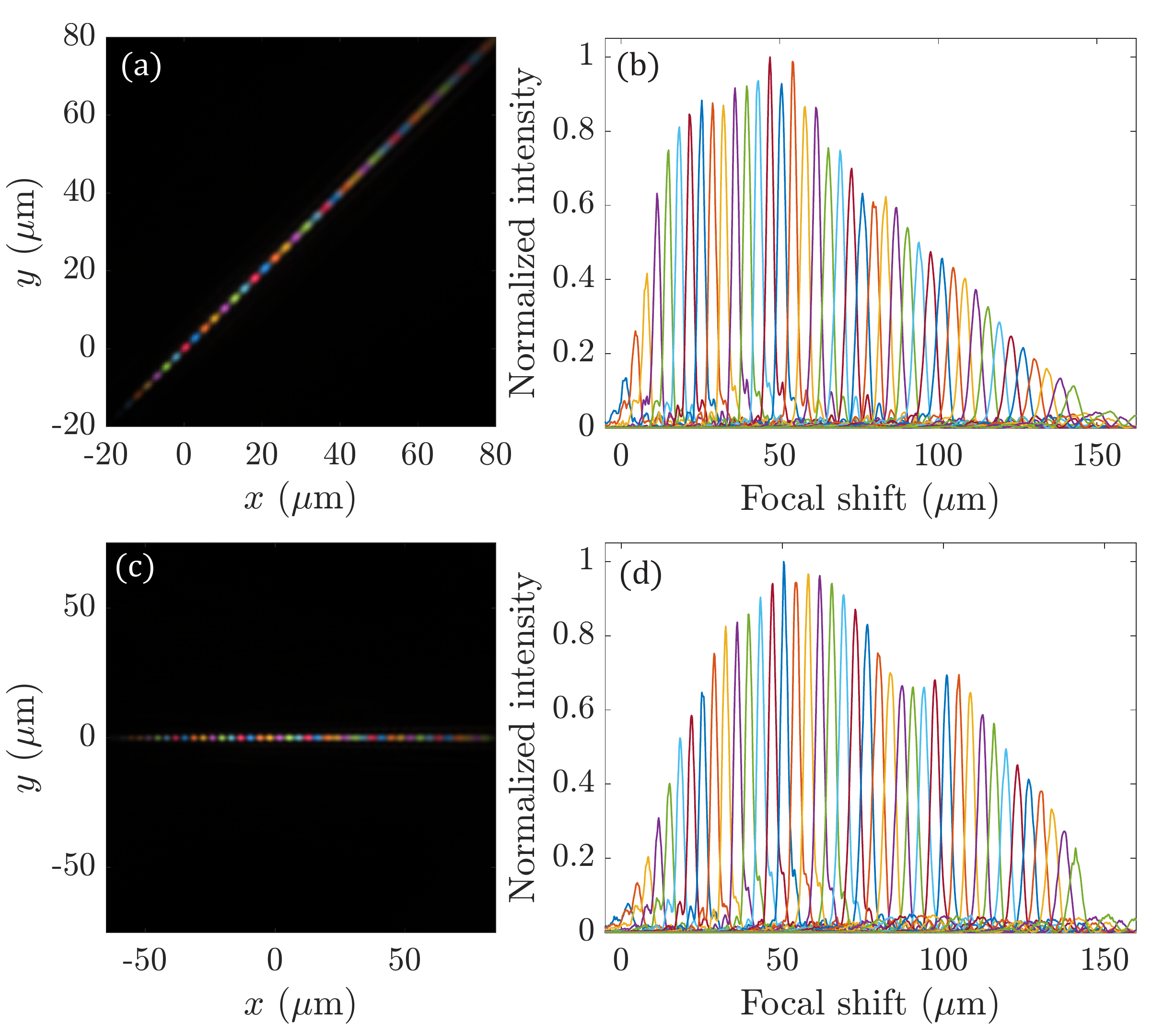}
    \caption{Simulated intensity profiles (false color image) of the different incident wavelengths focused on the $z=300\,\mu\mathrm{m}$ achromatic focal plane. Focal spots are distributed along the direction angles (a) $\alpha=45\degree$, (c) $\alpha=0\degree$. Also shown are simulated normalized intensity line-cuts for different wavelengths at the achromatic focal plane evaluated with (b) $\alpha=45\degree$ line-trajectory, (d) $\alpha=0\degree$ line-trajectory.}\label{fig: angular dispersion}
\end{figure}

During training, we set the object spectral resolution to be $\Delta\lambda=5~\mathrm{nm}$, which is significantly improved compared to what previously reported using ultracompact modulated axilenses \cite{Britton:2020_ChirpedAxilens,Chen:2020_Axilens_AO,Chen:2020_Axilens_OL}. We sampled $n=200$ incident wavelengths evenly spaced in the range from $\lambda_1=400~\mathrm{nm}$ to $\lambda_n=800~\mathrm{nm}$, which corresponds to the spectral range of the visible light. Given the focal position coordinates of the starting wavelength $\lambda_1$ denoted as $(x_1,y_1)$, the positions of the other wavelengths are determined using Eqs.\,\ref{eq: focal spot center x},\,\ref{eq: focal spot center y} for a given resolving power. The batch size is selected as $B=5$. We applied the uniform random initialization for the phase values of each pixel at the beginning of the training. The a-D$^2$NN is trained over $2000$ iterations using the Adam optimizer with a learning rate equals to $0.1$. The coefficients for updating adaptive weights are all $\gamma=1$. Our deep learning algorithm is developed within the flexible TensorFlow framework. The machine used for training consists of a GeForce RTX 3060 Ti graphical processing unit (GPU, Nvidia Inc.) with 8 GB memory, alongside with an Intel i9-10850K central processing unit (CPU, Intel Inc.) and 32 GB RAM. The typical training time for the SGD process is $\sim15$ minutes. 

At the end of the training we obtain a device with a two-layer phase modulated profile that achieves the desired spectral resolution $\Delta\lambda$ for a given pixel size $\Delta x$. Therefore, the inverse design of ultracompact spectrometers based on a-D$^2$NNs allows us to target different values of $\Delta\lambda$ independently of $\Delta x$, thus reducing the complexity of manufacturing process. Moreover, we can control the linear focusing trajectories along the horizontal and vertical directions, thus making the spectrometer capable of dispersing incident wavelengths onto any arbitrary two-dimensional path in the output plane. 

\begin{figure}[htbp]
    \centering\includegraphics[width=\linewidth]{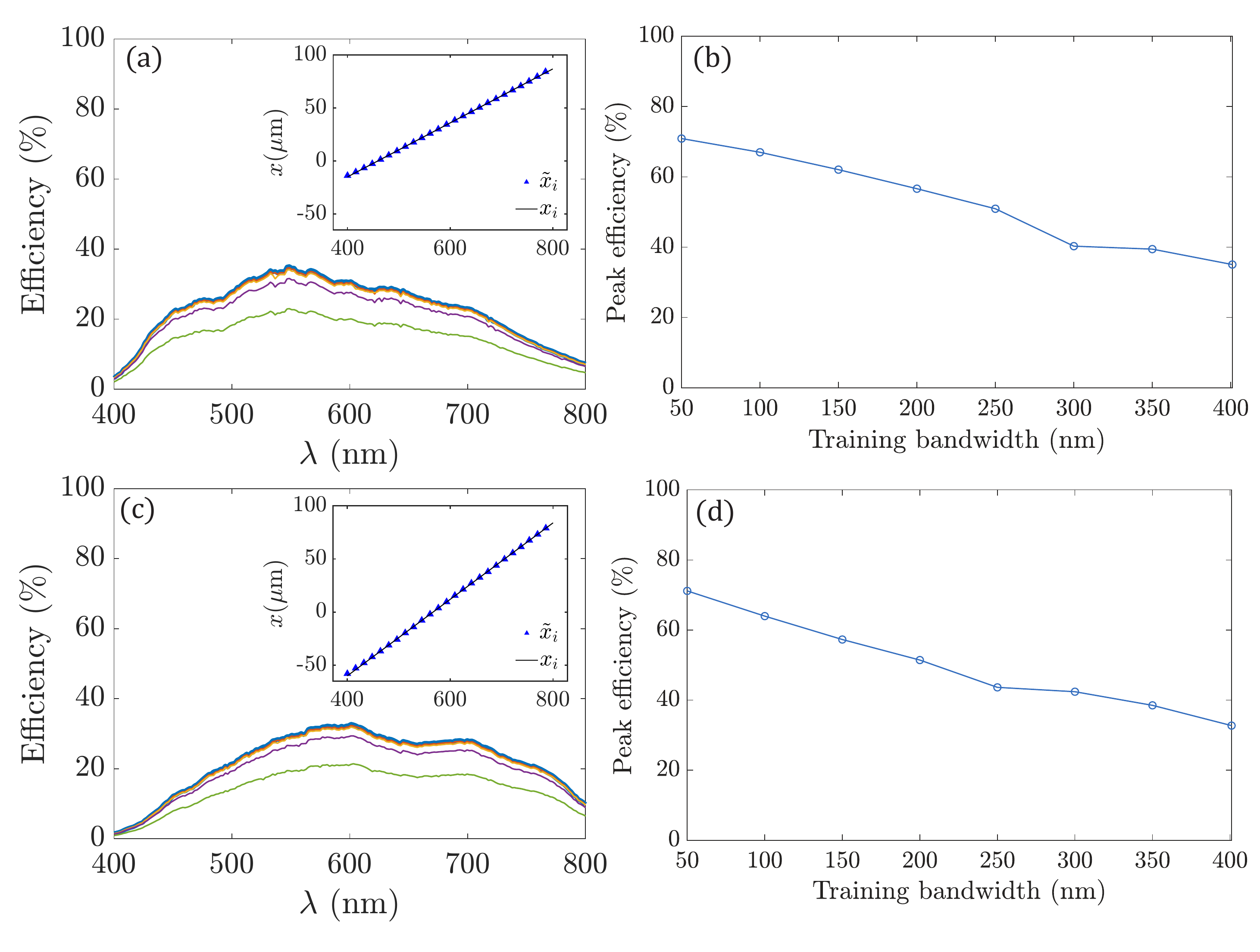}
    \caption{Focusing efficiency spectra for (a) $\alpha=45\degree$\,diagonal line, (c) $\alpha=0\degree$ horizontal line setup. Both figures compare the focusing efficiencies with respect to continuous phase (blue) and different discretized phase levels (orange for 32 levels, yellow for 16 levels, purple for 8 levels, and green for 4 levels). Corresponding insets show $x$ coordinates of simulated focal spots (blue triangle markers) and the target focal spots (black solid line). Note that the simulated focal spot positions agree well with predictions from Eqs.\,\ref{eq: focal spot center x},\,\ref{eq: focal spot center y}. Also shown are the peak efficiencies with respect to the training bandwidth of both setups in (b) and (d), the peak efficiency drops by almost a factor of 2 as the training bandwidth broadens from 50\,nm to 400\,nm. }\label{fig: focusing efficiency}
\end{figure}

As an example, in Fig.\,\ref{fig: angular dispersion} (a, c) we show the simulated focal plane intensity distributions along a $\alpha=45\degree$  diagonal line and a $\alpha=0\degree$ horizontal line respectively, while in Fig.\,\ref{fig: angular dispersion} (b, d) we show the line cuts of the intensity (normalized to its maximum value) along the corresponding directions. These incident wavelengths all separate from each other with a 5-$\mathrm{nm}$ spacing, which corresponds to the $\Delta\lambda$ used in our design. Note that we utilized false colors to better visualize the intensity distributions with respect to each wavelength. It can be clearly observed that the focal spots of the incident wavelengths are well-separated on the focal plane. These results demonstrate that our device can focus multiple incident wavelengths simultaneously and diffract them onto distinct transverse positions on the same achromatic plane. By setting different directions of the angular dispersion, we show the capability to map the wavelengths onto positions distributed on arbitrary trajectories. 

We then characterize the focusing behavior of the spectrometer device by first evaluating the simulated focusing efficiency spectrum $\hat{\eta}$ in Fig.\,\ref{fig: focusing efficiency} (a, c) with respect to the two angular dispersion directions introduced above. We note that $\hat{\eta}$ has broad spectrum with peak efficiency equals to approximately $40\%$. Practical fabrication technology, such as the one used in DOEs and metasurface engineering, requires the discretization of the phase profile into multiple levels. The impact of phase discretization of the device performances is shown by the curves with different colors in Fig.\,\ref{fig: focusing efficiency} (a, c), where we show the focusing efficiency spectra for different numbers of discrete phase levels. The results demonstrate that an $8$-level device already approaches the ideal performances of the one of the continuous phase. We also examined the dispersion behavior of the focal spots, which are located by finding the position of local maxima of each intensity distribution with respect to the incident wavelength. The insets in Figs.\,\ref{fig: focusing efficiency} (a, c) show the comparison between the focal spot positions simulated by forward propagating within a-D$^2$NN and those given by Eqs.\,\ref{eq: focal spot center x} and \ref{eq: focal spot center y}, demonstrating an excellent agreement. Next we investigated the effect of the training bandwidth on the focusing efficiency. As shown in Fig.\,\ref{fig: focusing efficiency} (b, d), when the training bandwidth broadens from $50\,\mathrm{nm}$ to $400\,\mathrm{nm}$, the peak efficiency drops by almost a factor of 2. Interestingly, we found that such a decrease is approximately linear when  increasing the training bandwidth. 
\begin{figure}[htbp]
    \centering\includegraphics[width=\linewidth]{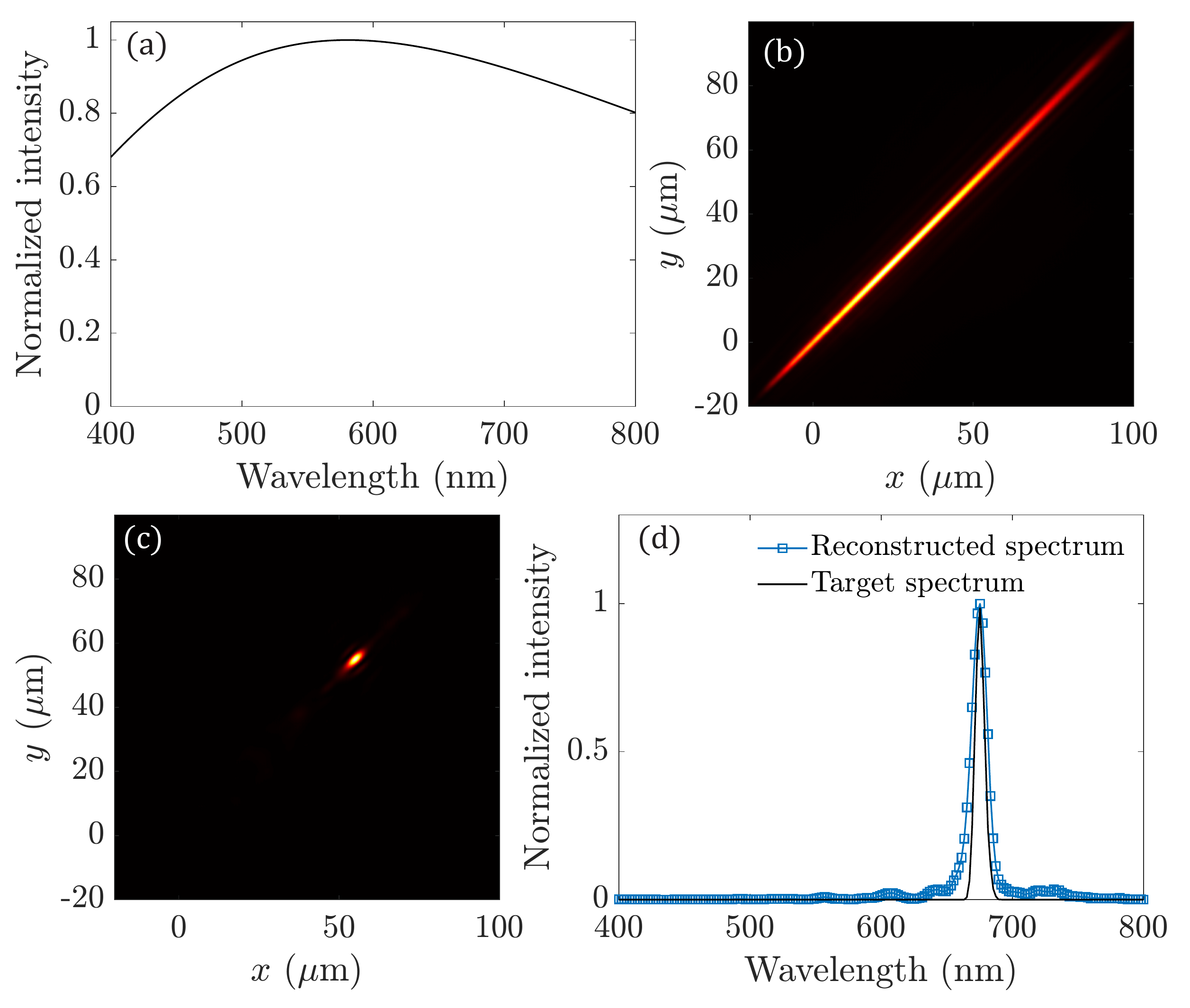}   
    \caption{(a) Simulated intensity distribution at $z=300\,\mu\mathrm{m}$ when the device is illuminated with a black body reference source. (b) Black body emission spectrum in $400$ to $800\,\mathrm{nm}$ range. (c) Simulated intensity distribution at the same achromatic plane when the device is illuminated by a superluminescent diode. (d) Reconstructed spectrum of the target diode source (blue square markers) and the target spectrum (black solid line). }\label{fig: reconstruction}
\end{figure}

Finally, we demonstrate that our multi-layer ultracompact spectrometer device can be used for the  single-shot spectral reconstruction of a practical light source. We first calibrated the spectral response of the device using the black body radiation as a reference source. The black body emission spectrum is given by Planck's law as: 
\begin{equation}\label{eq: black body radiation}
    S_{\mathrm{ref}}(\lambda,T)=\frac{2hc^3}{{\lambda}^5}\frac{1}{e^{\frac{hc}{kT\lambda}}-1} 
\end{equation}
where we considered $T=5000\,\mathrm{K}$ without loss of generality. The black body spectrum in the target spectral region is shown in Fig. \ref{fig: reconstruction} (a). After calibration, we simulated the spatial intensity distributions for both the black body reference and the target source (QSDM-680-2 superluminescent diode, QPhotonics LLC), which are shown in Fig. \ref{fig: reconstruction} (b) and (c), respectively. We denoted these intensity distributions as $I_\mathrm{ref}(x',y')$ and $I(x',y')$ respectively. Using Eqs.\,\ref{eq: focal spot center x},\,\ref{eq: focal spot center y} we then established a one-to-one mapping between the spatial distributions of the intensities and the corresponding spectral distributions (i.e., $I_\mathrm{ref}(\lambda)$ and $I(\lambda)$), for a given direction of the angular dispersion (i.e., $\alpha=45\degree$ in Fig.\,\ref{fig: reconstruction}) The emission spectrum of the diode is finally reconstructed via the following formula \cite{Britton:2020_ChirpedAxilens}: 
\begin{equation}\label{eq: spectrum reconstruction}
    S(\lambda)=I(\lambda)\frac{S_{\mathrm{ref}}(\lambda)}{I_\mathrm{ref}(\lambda)}
\end{equation}
The reconstruction of the diode spectrum obtained from Eq.\,\ref{eq: spectrum reconstruction} matches very well the target curve in Fig.\,\ref{fig: reconstruction} (d). 

To conclude, we proposed and developed an inverse design approach based on a-D$^2$NNs for ultracompact spectrometers trained to maximize the focusing efficiency over a broad band of wavelengths with customized focal trajectories. In particular, we demonstrated broad focusing efficiency spectra peaked at around $40\%$ across the visible and with a spectral resolution $\Delta\lambda=5~\mathrm{nm}$ for devices with $L=100~\,\mu\mathrm{m}$ side length. We also established the fundamental trade-off between spectral bandwidth and focusing efficiency. Moreover, the proposed concepts can naturally be extended to spectral bands other than visible. The flexible a-D$^2$NN approach introduced here for the design of ultracompact focusing spectrometers enables novel broadband diffractive devices with desired angular and spatial dispersion behavior for applications to multispectral imaging, multi-band detection, and lensless microscopy.

\begin{backmatter}
\bmsection{Funding}
National Science Foundation (ECCS-2015700).

\bmsection{Disclosures}
The authors declare no conflicts of interest.

\bmsection{Data Availability Statement}
Data underlying the results presented in this paper are
not publicly available at this time but may be obtained from the authors upon
reasonable request.

\end{backmatter}

\bibliography{ref_list}

\begin{thebibliography}{10}
\newcommand{\enquote}[1]{``#1''}

\bibitem{Demtroeder:2014_Laser}
W.~Demtr{\"o}der, \emph{Laser spectroscopy 1: basic principles} (Springer,
  2014).

\bibitem{He:1998_EDG_AWG}
J.-J. He, B.~Lamontagne, A.~ge, L.~Erickson, M.~Davies, and E.~S. Koteles,
  {\protect\JournalTitle{J. Lightwave Technol.}} \textbf{16}, 631 (1998).

\bibitem{Janz:2004_EDG_AWG}
S.~Janz, A.~Balakrishnan, S.~Charbonneau, P.~Cheben, M.~Cloutier,
  A.~Del\^{a}ge, K.~Dossou, L.~Erickson, M.~Gao, P.~Krug, B.~Lamontagne,
  M.~Packirisamy, M.~Pearson, and D.-X. Xu, {\protect\JournalTitle{IEEE
  Photonics Technology Letters}} \textbf{16}, 503 (2004).

\bibitem{Ma:2013_EDG}
X.~Ma, M.~Li, and J.-J. He, {\protect\JournalTitle{IEEE Photonics Journal}}
  \textbf{5}, 6600807 (2013).

\bibitem{Babin:2009_Holo}
S.~Babin, A.~Bugrov, S.~Cabrini, S.~Dhuey, A.~Goltsov, I.~Ivonin, E.-B. Kley,
  C.~Peroz, H.~Schmidt, and V.~Yankov, {\protect\JournalTitle{Applied Physics
  Letters}} \textbf{95}, 041105 (2009).

\bibitem{Peroz:2012_Holo}
C.~Peroz, C.~Calo, A.~Goltsov, S.~Dhuey, A.~Koshelev, P.~Sasorov, I.~Ivonin,
  S.~Babin, S.~Cabrini, and V.~Yankov, {\protect\JournalTitle{Opt. Lett.}}
  \textbf{37}, 695 (2012).

\bibitem{Calafiore:2014_Holo}
G.~Calafiore, A.~Koshelev, S.~Dhuey, A.~Goltsov, P.~Sasorov, S.~Babin,
  V.~Yankov, S.~Cabrini, and C.~Peroz, {\protect\JournalTitle{Light: Science
  {\&} Applications}} \textbf{3}, e203 (2014).

\bibitem{Cheben:2007_AWG}
P.~Cheben, J.~H. Schmid, A.~Del\^{a}ge, A.~Densmore, S.~Janz, B.~Lamontagne,
  J.~Lapointe, E.~Post, P.~Waldron, and D.-X. Xu, {\protect\JournalTitle{Opt.
  Express}} \textbf{15}, 2299 (2007).

\bibitem{Zou:2020_AWG}
J.~Zou, X.~Ma, X.~Xia, J.~Hu, C.~Wang, M.~Zhang, T.~Lang, and J.-J. He,
  {\protect\JournalTitle{J. Lightwave Technol.}} \textbf{38}, 4447 (2020).

\bibitem{Zhu:2017_ChirpedMetalens}
A.~Y. Zhu, W.-T. Chen, M.~Khorasaninejad, J.~Oh, A.~Zaidi, I.~Mishra, R.~C.
  Devlin, and F.~Capasso, {\protect\JournalTitle{APL Photonics}} \textbf{2},
  036103 (2017).

\bibitem{Britton:2020_ChirpedAxilens}
W.~A. Britton, Y.~Chen, F.~Sgrignuoli, and L.~Dal~Negro,
  {\protect\JournalTitle{ACS Photonics}} \textbf{7}, 2731 (2020).

\bibitem{Chen:2020_Axilens_AO}
Y.~Chen, W.~A. Britton, and L.~Dal~Negro, {\protect\JournalTitle{Appl. Opt.}}
  \textbf{59}, 5532 (2020).

\bibitem{Chen:2020_Axilens_OL}
Y.~Chen, W.~A. Britton, and L.~Dal~Negro, {\protect\JournalTitle{Opt. Lett.}}
  \textbf{45}, 2371 (2020).

\bibitem{Britton:2021_DualBand_LPR}
W.~A. Britton, Y.~Chen, F.~Sgrignuoli, and L.~Dal~Negro,
  {\protect\JournalTitle{Laser \& Photonics Reviews}} \textbf{15}, 2000207
  (2021).

\bibitem{Khorasaninejad:2016_MetalensSpec}
M.~Khorasaninejad, W.~T. Chen, J.~Oh, and F.~Capasso,
  {\protect\JournalTitle{Nano Letters}} \textbf{16}, 3732 (2016). PMID:
  27119987.

\bibitem{Redding:2013_Random}
B.~Redding, S.~F. Liew, R.~Sarma, and H.~Cao, {\protect\JournalTitle{Nature
  Photonics}} \textbf{7}, 746 (2013).

\bibitem{Rahnama:2020_ChirpedFilamentArrayGrating}
A.~Rahnama, K.~Mahmoud~Aghdami, Y.~H. Kim, and P.~R. Herman,
  {\protect\JournalTitle{Advanced Photonics Research}} \textbf{1}, 2000026
  (2020).

\bibitem{Emadi:2012_LVOF}
A.~Emadi, H.~Wu, G.~de~Graaf, and R.~Wolffenbuttel, {\protect\JournalTitle{Opt.
  Express}} \textbf{20}, 489 (2012).

\bibitem{Li:2021_SWF}
A.~Li and Y.~Fainman, {\protect\JournalTitle{Nature Communications}}
  \textbf{12}, 2704 (2021).

\bibitem{Sharkawy:2001_ResPhotonicDefect}
A.~Sharkawy, S.~Shi, and D.~W. Prather, {\protect\JournalTitle{Appl. Opt.}}
  \textbf{40}, 2247 (2001).

\bibitem{Nitkowski:2008_ResMicroring}
A.~Nitkowski, L.~Chen, and M.~Lipson, {\protect\JournalTitle{Opt. Express}}
  \textbf{16}, 11930 (2008).

\bibitem{Xia:2011_ResMicrodonut}
Z.~Xia, A.~A. Eftekhar, M.~Soltani, B.~Momeni, Q.~Li, M.~Chamanzar,
  S.~Yegnanarayanan, and A.~Adibi, {\protect\JournalTitle{Opt. Express}}
  \textbf{19}, 12356 (2011).

\bibitem{Liu:2018_MLInverseDesign}
D.~Liu, Y.~Tan, E.~Khoram, and Z.~Yu, {\protect\JournalTitle{ACS Photonics}}
  \textbf{5}, 1365 (2018).

\bibitem{Ma:2021_MLInverseDesignReview}
W.~Ma, Z.~Liu, Z.~A. Kudyshev, A.~Boltasseva, W.~Cai, and Y.~Liu,
  {\protect\JournalTitle{Nature Photonics}} \textbf{15}, 77 (2021).

\bibitem{Liu:2021_MLInverseDesign}
Z.~Liu, D.~Zhu, L.~Raju, and W.~Cai, {\protect\JournalTitle{Advanced Science}}
  \textbf{8}, 2002923 (2021).

\bibitem{Lin:2018_D2NN}
X.~Lin, Y.~Rivenson, N.~T. Yardimci, M.~Veli, Y.~Luo, M.~Jarrahi, and A.~Ozcan,
  {\protect\JournalTitle{Science}} \textbf{361}, 1004 (2018).

\bibitem{Luo:2019_D2NNFilter}
Y.~Luo, D.~Mengu, N.~T. Yardimci, Y.~Rivenson, M.~Veli, M.~Jarrahi, and
  A.~Ozcan, {\protect\JournalTitle{Light: Science {\&} Applications}}
  \textbf{8}, 112 (2019).

\bibitem{Veli:2021_D2NNTerahertz}
M.~Veli, D.~Mengu, N.~T. Yardimci, Y.~Luo, J.~Li, Y.~Rivenson, M.~Jarrahi, and
  A.~Ozcan, {\protect\JournalTitle{Nature Communications}} \textbf{12}, 37
  (2021).

\bibitem{Chen:2022_D2NNDualLens}
Y.~Chen, Y.~Zhu, W.~A. Britton, and L.~Dal~Negro, {\protect\JournalTitle{Opt.
  Lett.}} \textbf{47}, 2842 (2022).

\bibitem{Banerji:2019_ReviewFlatOptics}
S.~Banerji, M.~Meem, A.~Majumder, F.~G. Vasquez, B.~Sensale-Rodriguez, and
  R.~Menon, {\protect\JournalTitle{Optica}} \textbf{6}, 805 (2019).

\bibitem{Lalanne:2017_MetalensReview}
P.~Lalanne and P.~Chavel, {\protect\JournalTitle{Laser \& Photonics Reviews}}
  \textbf{11}, 1600295 (2017).

\bibitem{Khorasaninejad:2016_Science}
M.~Khorasaninejad, W.~T. Chen, R.~C. Devlin, J.~Oh, A.~Y. Zhu, and F.~Capasso,
  {\protect\JournalTitle{Science}} \textbf{352}, 1190 (2016).

\bibitem{Yilmaz:2019_Metasurface}
N.~Yilmaz, A.~Ozdemir, A.~Ozer, and H.~Kurt, {\protect\JournalTitle{Journal of
  Optics}} \textbf{21}, 045105 (2019).

\bibitem{Voelz:2011_MATLAB}
D.~G. Voelz, \emph{Computational fourier optics: a MATLAB tutorial}, vol. 534
  (SPIE press Bellingham, Washington, 2011).

\end{thebibliography}

\bibliographyfullrefs{ref_list}


\ifthenelse{\equal{\journalref}{aop}}{%
\section*{Author Biographies}
\begingroup
\setlength\intextsep{0pt}
\begin{minipage}[t][6.3cm][t]{1.0\textwidth} 
  \begin{wrapfigure}{L}{0.25\textwidth}
    \includegraphics[width=0.25\textwidth]{john_smith.eps}
  \end{wrapfigure}
  \noindent
  {\bfseries John Smith} received his BSc (Mathematics) in 2000 from The University of Maryland. His research interests include lasers and optics.
\end{minipage}
\begin{minipage}{1.0\textwidth}
  \begin{wrapfigure}{L}{0.25\textwidth}
    \includegraphics[width=0.25\textwidth]{alice_smith.eps}
  \end{wrapfigure}
  \noindent
  {\bfseries Alice Smith} also received her BSc (Mathematics) in 2000 from The University of Maryland. Her research interests also include lasers and optics.
\end{minipage}
\endgroup
}{}

\end{document}